# Teachers are from heaven, students are from hell – true or false?

## Luiz Fernando Capretz

**Abstract:** *Researchers have long tried to relate personality types to teaching and learning styles. It is believed that the psychological theory behind the Myers–Briggs Type Indicator (MBTI) can help university teachers to accept variety in teaching and learning approaches. This paper makes some assertions about the personality traits of academics and students. These traits can create harmony or discord for individual students, depending on whether their approach to learning matches the teacher's approach to teaching. Although some teaching strategies can be useful for a whole class, differences among students make it necessary to diversify those strategies.*

**Keywords:** *teaching style; learning style; higher education; personality types; MBTI*

The author is with Department of Electrical and Computer Engineering, University of Western Ontario, London, Ontario N6G 1H1, Canada. E-mail: lcapretz@eng.uwo.ca.

*'The meeting of two personalities is like the contact of two chemical substances:*

> *if there is any reaction, both are transformed.'*

> *(Carl G. Jung)*

## Background

Each of us has something to learn and to teach. Adjusting instruction to accommodate the learning styles of different types of students can increase both the students' achievement and their enjoyment of learning. Many teachers believe that 'being fair' means treating all students equally. However, if this translates into using the same approach for every student, or treating all students in the same way, then problems are likely to arise for those students who feel left out because a teacher chooses classroom activities that reflect his or her own teaching style. Once the natural and healthy differences that exist in students are fully understood, teachers will appreciate that 'being fair' really means providing equal opportunities for each student to learn in the manner that best suits his or her own natural way of learning. This approach can improve the degree of satisfaction and understanding among both university teachers and students.

Educators have been using the Myers–Briggs Type Indicator (MBTI) (Myers *et al*, 1998) to develop teaching methods and to understand individual learning styles and differences in motivation. In this study, MBTI is used not only to categorize students but also to expose their learning differences, strengths and weaknesses. The intention is to apply this well-researched view of personality type to a discussion of teaching and learning styles. Thus several approaches to teaching, and how personality type is related to each approach, are discussed. It is suggested that this is the best way to improve teaching effectiveness, because it explains why teachers sometimes pressure themselves to teach in a way that does not suit their personality styles and why students are forced to learn in environments that do not always suit their ways of learning. To understand this, it is necessary to look at a teacher's and a student's preferred teaching and learning styles, as will be discussed later in this article.

The MBTI is an instrument designed to measure four dimensions of an individual's personality: extroversion/introversion, sensing/intuitive, thinking/feeling and judging/perceiving. As a starting point, therefore, it is useful briefly to introduce the four scales of the MBTI. The indicator establishes four parameters for assessing personality types. We all have personality qualities of each scale or parameter; we simply prefer some qualities or are more comfortable with some styles than others, just as right-handers are more comfortable using the right hand, but sometimes may use the left hand. Each respondent is forced to choose preferences; the higher the score on each preference, the stronger the preference is likely to be.

*Extroversion (E) and Introversion (I)*

The first scale represents complementary attitudes towards the outer world of people and action; or the inner world of ideas and private things. The extrovert prefers to look outward, while the introvert looks inward. For example, strong extroverts (*E*) are sometimes said to 'talk to think', whereas the introverts 'think to talk'. The implications of these terms go beyond the everyday caricatures of sociable versus shy. Extroverts are talkative, initiators of conversations and are outgoing: they like action and variety. In contrast, introverts (*I*) are quiet, respondents in conversation and are reserved, liking silence and time to consider things. To an extrovert, the outside world is where interesting things happen, while for an introvert the hustle and bustle of the outside world can be a distraction from the more interesting world of thoughts and ideas. Both orientations can be valuable, just as both can be costly, but neither is inherently better than the other: they are simply different.

*Sensing (S) and Intuition (N)*

The second scale of preference distinguishes the way we take in information from the environment. Whereas a strong *S* type might need to assimilate a whole series of facts in linear fashion, the person who prefers *N* can absorb the same information through abstraction and concepts that may not at first seem directly related, but that could establish a pattern. An *S* person enjoys using skills already learned more than learning new ones, and dislikes new problems unless prior experience shows how to solve them. *N* likes to use new skills rather than practise old ones, and dislikes doing the same thing over and over again. This scale indicates whether a person would rather understand the objects, events and details of the present (*S*) or imagine the possibilities of the future (*N*). The adjectives that describe a sensing person are: realistic, practical and fact-oriented. Those appropriate to an intuitive person are: speculative, imaginative and creative. Of course, we all share both sets of qualities to some extent, but one set predominates.

*Thinking (T) and Feeling (F)*

The third mode of orientation in the MBTI classification is thinking and feeling; again these terms are more extensive than everyday usage indicates. This scale shows a person's preferred basis for making decisions: logical analysis (*T*) or personal values (*F*). This scale of preferences identifies thinking as the analytical way of making a decision, while feeling describes the tendency to rely on values to make decisions. Whereas the *T* individual needs distance from a situation to make a decision, the *F* person must be immersed in the situation in order to gain empathy with the people involved before making a decision. A *T* individual may neglect and hurt other people's feelings without knowing it; that rarely happens to an *F* person, who is usually very aware of other people's feelings. Thinking people are principle-oriented, cool-headed and firm. Feeling people are emotion-oriented, warm-hearted and gentle. There is a gender difference in the general population regarding this scale: that is, the majority of women prefer feeling (*F*), but the majority of men prefer thinking (*T*).

*Judging (J) and Perceiving (P)*

The fourth scale differentiates between how we are oriented in our lifestyles, and how we organize our world. It reveals whether a person favours organizing and controlling events (*J*) or observing and adapting to them (*P*). *J* identifies the tendency to be a super-organized character. If a deadline is to be met, the *J* person usually finishes the task well in advance. At the other extreme the person who prefers perceiving (*P*) appears to be very disorganized and seems to be distracted from completing a task until some little bell goes off at the last minute and tells this individual to get it done. It is often said that the easiest way to distinguish between these two preferences is to look at the person's desk. The desk of a *J* person is immaculately organized; the desk of a *P* individual appears to be in constant chaos even though the *P* person claims to know exactly where everything is located and that there are rules underlying the chaos. The words deadlines, punctual, schedule and routine apply to judging types, whereas open-ended, flexible, adaptable and spontaneous apply more to perceiving types.

In simple words, the MBTI filters these four scales of preferences (each scale has a pair of opposite traits) to sort out a person's preferred type. Hence, there are 16 possible configurations, as shown in Table 1. If the MBTI results show that a person is *ISTP* (introvert, sensing, thinking, and perceiving), then the terminology is to suggest that the person *prefers ISTP*, not that the person *is* an *ISTP*. Again, people can and do use all eight preferences. In each of the four dimensions, however, we all have one preference that is stronger than the other, one that works better for us than its complement.

**Table 1. The 16 MBTI types.**

| | | | |
|------|------|------|------|
| ISTJ | ISFJ | INFJ | INTJ |
| ISTP | ISFP | INFP | INTP |
| ESTP | ESFP | ENFP | ENTP |
| ESTJ | ESFJ | ENFJ | ENTJ |

During the past decade, MBTI has received considerable attention and use in a variety of applied settings. Cooper and Miller (1991) report that the level of learning style/teaching style congruency is related to academic performance and to student evaluations of the course and instructor. Moreover, the discrepancy between students' preferences for learning in a concrete manner (*S* type) and the academic's penchant for teaching in abstractions (*N* type) appears to contribute to student dissatisfaction, as indicated by course and instructor evaluations.

In addition, Gardner and Jewler (1992) suggest that college students can improve their study habits by knowing their MBTI type, and they show that different learning styles are associated with each preference; advice is also provided for the student whose learning style conflicts with the instructor's teaching style. However, Pittinger (1993) claims that the MBTI makes few unique practical or theoretical contributions to the understanding of behaviour; motivation and learning are more complex than any theory intended to enlighten our practice. Thus a word of caution is necessary: a knowledge of learning preferences must be used with the understanding that we cannot fully explain all aspects of human behaviour. When it comes to people, there are few simple answers.

The learning preferences described in this work can help teachers understand how their students learn. They are applicable to all students to some extent; however, they apply more clearly to some than to others. Therefore, teachers should use the learning preferences when they are helpful, and not as categories into which they expect all students to fit.

## How types teach

We tend to teach as we ourselves like to be taught. Commonly we assume that our students can learn best by employing the same techniques that we used as students. However, people differ significantly in the ways they learn best, and it is believed that these learning styles are connected with personality types. Following the MBTI theory, Table 2 relates some aspects of personality traits to teaching (Fairhurst and Fairhurst, 1995).

We would certainly not suggest that instructors should always adapt to the learning styles of their students. This is not only impossible in a diverse classroom setting, but it also creates too much stress on the instructor. Certainly, an instructor can modify an approach for those students who may feel disconnected. Consider the following scenario. Extroverted instructors tend to be more activity-oriented, while introverted instructors usually like to allow more time for reflection. An instructor who makes use of a lot of discussion in the classroom, for example, could be aware of the difficulty that introverts might have with the approach, and be supportive rather than punitive when introverted students are slow to become involved in debates. Instructors can also use private tutorials and other opportunities to individualize their instruction (that is, to teach in a way suitable to a particular student).

**Table 2. Types and teaching styles.**

| | |
|---|---|
| **Extroverts (*E*)** <br> *E* teachers give students choices and voice, are attuned to changes in students' attention and comfortable with noisy classrooms. *E*s tend to evaluate enthusiastic and talkative students positively. | **Introverts (*I*)** <br> *I* teachers structure teaching activities, are attuned to the ideas they teach and are comfortable with a silent classroom. *I*s tend to evaluate thoughtful and reflective students positively. |
| **Sensing (*S*)** <br> S instructors emphasize practical information and concrete skills and usually ask detailed and fact-oriented questions. *S*s are biased towards students who are practical and accurate. | **Intuitive (*N*)** <br> *N* instructors emphasize concepts and the implications of facts and their questions call for synthesis and meaning. *N*s are biased towards students who are conceptual and insightful. |
| **Thinking (*T*)** <br> *T* professors talk from an objective base. They want students to focus on what they doing or saying; they attend to the class as a whole. *T*s are inclined towards students who are logical, precise and critical of their own work. | **Feeling (*F*)** <br> *F* professors seek dialogue and engagement. They encourage students to focus on interpersonal work and attend to individuals or small groups. *F*s are inclined towards students who are personable and friendly. |
| **Judging (*J*)** <br> *J* teachers are very orderly and stick to a class plan with organized lectures, they like a well-arranged classroom. *J*s tend to evaluate task-focused, punctual and systematic students positively. | **Perceiving (*P*)** <br> *P* teachers are more casual and less organized. They like to handle as man acivities as possible. *P*s tend to evaluate spontaneous, adaptable and easygoing students positively. |

Additionally, effective teaching is achieved by combining explanation of basic principles and their meanings with concrete facts and examples. That means that a student will learn best when a general description of the basic idea is supported by the kinds of examples that will lead to the student's spontaneous understanding. This is especially true for the sensing student. Effective teaching is also significantly enhanced by the emotional strength of the instructor capable of captivating the feeling student. Therefore, teaching is most effective if different but complementary styles are applied and combined; the integration of different techniques avoids burn-out and boredom. The ideal instructor, then, is one who can select from an armoury of skills and techniques the appropriate strategy for enhancing learning.

## How types learn

As already noted, the MBTI types indicate the student's preference on four scales; each scale has two opposites. Remember that the student is capable of operating with either of these opposites, but that one side of the scale probably comes more naturally and requires less effort, like using the preferred hand. Table 3 contains a summary of findings that relate personality type to ways of learning, as described by DiTiberio and Hammer (1993). For example, the table indicates how the process of learning is fundamentally different for sensing and intuitive students. Therefore, if a student is having difficulty learning new material it may be because he or she is trying to learn in a way that is not consistent with his or her natural style.

Extroverted students usually learn best in an active environment, and have trouble sitting for long periods of time listening to a lecture or writing a paper. They often work best when they can interact in small groups or talk lessons over with a partner. These students tend to plunge into activities without much forethought, relying on trial-and-error rather than anticipation to solve problems; they like to speak their thoughts as they occur. On the other hand, introverted students usually learn best when they work quietly and alone, reading lessons over or writing them out before discussion. They like to think through a problem before talking about it. Introverted students should be given adequate time to formulate their responses before discussing the problem, and are more comfortable when they can prepare their responses in advance, as they like to keep thoughts inside until they are polished.

**Table 3. Learning styles associated with each preference.**

| Extroverts (*E*) | Introverts (*I*) |
|---|---|
| Learn best when in action | Learn best by pausing to think |
| Like to study with others | Prefer to study individually |
| Find background sounds help them study | Need quiet for concentration |
| Want teachers who encourage class discussion | Want teachers who give clear lectures |

| Sensing (*S*) | Intuition (*N*) |
|---|---|
| Memorize facts | Use imagination to go beyond facts |
| Value what is practical | Value what is original |
| Like hands-on experience | Like theories to give perspective |
| Want teachers who give clear assignments | Want teachers who enforce independent thinking |

| Thinking (*T*) | Feeling (*F*) |
|---|---|
| Want objective material to study | Want to relate to the material personally |
| Use logic to guide learning | See personal values as important |
| Learn by challenge and debate | Learn by being supported and appreciated |
| Want teachers who make logical presentations | Want teachers who establish personal rapport |

| Judging (*J*) | Perceiving (*P*) |
|---|---|
| Like to have things settled | Value change and adapt to events |
| Plan work well in advance | Keep their options open |
| Work steadily toward goals | Work impulsively with bursts of energy |
| Want teachers to be business-like | Want teachers to be entertaining |

Sensing students prefer the concrete to the abstract and tend to learn best in step-by-step progression. They follow clear, specific instruction and are often frustrated when given vague directions or unclear assignments; they are usually better at summarizing material than analysing it. They like films, demonstrations and audiovisuals, and prefer practical examples and hands-on exercises, as this requires actively engaging the senses. Intuitive students prefer the abstract to the concrete and can become bored during drill or factual lectures. They thrive in classroom situations that place a premium on imagination, but are sometimes careless about details. They welcome opportunities for brainstorming and can see the big picture. They work best if they can see global patterns, incorporate new approaches and demonstrate originality.

Thinking students prefer classrooms in which instructors provide a clear rationale for assignments.

They like topics that help them understand systems or cause-and-effect relationships, those which develop logical criteria. These students tend to think syllogistically and analytically; they work best if they can prepare outlines and state the objective first. On the other hand, feeling students prefer assignments in which they can find a human angle or have emotional investments. They are less concerned with logic than with values, and they like a situation in which helping people is the main activity. They see competition as disharmonious, and like instruction that involves feeling.

Judging students tend to seek closure. They are comfortable making decisions, and once a decision is made they stick to it. They tend to be well-organized, meet deadlines and usually prefer to work on one task at a time. They thrive in a structured classroom with systematically organized lectures and exercises; they like to follow a study schedule. Exactly the opposite, perceiving students tend to resist closure. They prefer spontaneity so that they can explore things without pre-planning. They like to undertake several tasks simultaneously and often work right up to and even beyond deadlines; they work best if they have independence and autonomy to complete the tasks.

To summarize, 'learning style' refers to an individual's characteristic and consistent approach to perceiving, organizing and processing information. The idea that people have different learning styles is enticing to educators. First, it highlights the importance of learning processes. Secondly, it is an egalitarian concept because it focuses on people's strengths and weaknesses; that is, learners become *different* rather than bad, poor, average, good and excellent. Because of this, it would be naïve to expect that instructors could easily design and deliver a course to fit the learning style needs of all their students.

## Conclusions

The MBTI is neither a measure of teaching performance nor learning competence; it is only an indicator of preferences. This limitation, however, does not preclude the possibility of using MBTI to improve higher education practices. As a rule of thumb, MBTI provides insights for effective teaching and learning, and it can be usefully employed as a guide for understanding learning styles and improving teaching skills.

We assume that learning style is stable during adulthood, as the MBTI is relatively constant. As the issue of the stability of learning preferences is likely to persist, it has implications for how best to provide advice and guidance to mature learners. If preferences are fixed, then it is better to attempt to match teaching and learning styles, develop alternative learning activities for people with different styles and guide people into those styles to which they are best suited. The implication for teachers is that they should be aware of a learner's style and apply corrective intervention where appropriate.

On the other hand, intervention is also possible by assisting students to diversify their learning strategies and encouraging optional choices outside a preferred style. Fortunately, almost every class has some students of each type, so by accommodating every preference each student can be exposed to other preferences. This helps students to develop skills in their non-preferred areas. Such development is beneficial in creating the balance that adults need to function effectively in the world.

A few professors may conclude that the need for an appropriate match between student type and teaching style is more critical for students of below-average intelligence. Others believe that university students should be sufficiently mature to accommodate whatever style a professor brings to the lecture. But this author believes that all students can benefit tremendously if there is matching of learning and teaching styles. If professors are careful to reject the idea the attitude 'I teach this way because I am who I am', their students can only benefit.

According to the MBTI model, people are at their best when they have natural command of their preferred function. Therefore, students learn most effectively, especially when approaching new or difficult topics, when they are given opportunities to use their natural way of learning. It seems reasonable to expect that students can encompass a variety of personality traits. Regarding learning styles, there is no one best combination of characteristics, since each preference has its own advantages and drawbacks. It is therefore a fallacy to think that professors can devise a single teaching technique that will always appeal to all students.

Rosati (1999) explores students' performances in an experiment concerning teaching presentation mode (one sensing and the other intuitive) and students' learning styles in problem-solving courses. The psychological theory behind the MBTI suggests that sensing types, in their learning, rely on experience rather than theory and have a preference for progressing from the familiar in a step-by-step manner. Intuitive types, on the contrary, rely more on inspiration, a fact which often leads to an ability to understand abstract, symbolic and theoretical relationships. Therefore, the MBTI intuitive/sensing dimension can separate the intuitive student with a preference for the abstract, global and theoretical from the sensing student with a preference for the practical, factual and specific.

The ability to think in the abstract is no better than the ability to think concretely. Each learning style has its strengths and weaknesses and therefore a person locked exclusively into one style is never going to be an ideal learner. In fact, he or she will just be an ordinary learner: educators should bear in mind that everyone has a learning style that narrows his or her capacity as a learner. This does not mean, however, that there are two classes of learners – a privileged class (learners who can overcome their limitations) and a less privileged class (learners who are incapable of using different learning styles). It is simply a matter of preference, of being more or less comfortable with a particular style. This challenges the notion that learning potential is reducible to a single dimension, such as intelligence.

Becoming a perfect learner entails integrating the bipolar dimensions of each personality type and ultimately of each learning style, and operating satisfactorily with any learning style, although there is a dominant style with which a student is more comfortable. By the same token, good professors should be able to broaden their repertoire of effective teaching techniques, and so to reach all students at least some of the time. They should also consider varying their teaching styles to motivate and establish common ground with those students who have traits different from their own.

Inspired by the MBTI, the author has developed a range of practices for effective teaching and learning in a software engineering course (Capretz, 2002). The aim is to reach every student, in different ways, by devising various teaching approaches and activities during a course. As software engineering teachers tend to be *ST* and *NT*, and software engineering students are also usually *ST* and *NT* (Capretz, 2003), teachers are already reaching out to the majority of their students. However, the teaching of software engineering courses would be more effective for other types, such as *SF*s, *NF*s and *E*s, if they were to incorporate and emphasize more open discussions and human factor issues. For instance, feeling types like to see the personal implications of a concept; in a software engineering course this can be achieved by a discussion on ethics, and the human side of software management and team interaction.

In summary, Tables 2 and 3 together set out some positive aspects of effective teaching and learning. When teachers and students are aware of their personality types, and consequently of their preferred ways of teaching and learning:

*   they view one another more positively;
*   students' expectations are matched, creating an optimal atmosphere for learning;
*   teachers evaluate the performance and intellects of students more accurately; and
*   the goal of their interaction, through the use of similar communication modes, is more likely to be achieved.

The answer to the question posed in the title of this article should, of course, be 'false'.